\documentclass[3p,times,procedia]{elsarticle}
\usepackage{nupha_ecrc}

\volume{00}

\firstpage{1}

\journalname{Nuclear Physics A}

\runauth{C. N. Cruz-Camacho et al.}

\jid{nupha}

\jnltitlelogo{Nuclear Physics A}

\usepackage{amstext,amssymb,amsthm,amscd,amsmath,amsfonts,mathtools,amsfonts}

\usepackage{caption}
\usepackage{subcaption}

\usepackage[figuresright]{rotating}

\newcommand{\besp}{\begin{split}}
\newcommand{\eesp}{\end{split}}
\newcommand{\x}{\hat{x}^\mu}

\newcommand{\ds}{dS_3\otimes R}

\newcommand{\tem}{\hat{T}}

\newcommand{\be}{\begin{equation}}
\newcommand{\ee}{\end{equation}}
\newcommand{\bea}{\begin{eqnarray}}
\newcommand{\eea}{\end{eqnarray}}
\newcommand{\bs}{\begin{subequations}}
\newcommand{\es}{\end{subequations}}
\newcommand{\beal}{\begin{align}}

\begin{document}

\begin{frontmatter}



\dochead{XXVIIth International Conference on Ultrarelativistic Nucleus-Nucleus Collisions\\ (Quark Matter 2018)}

\title{Out-of-equilibrium Gubser flow attractors}


\author[a]{C. N. Cruz-Camacho}
\author[b]{M. Martinez}

\address[a]{Departamento de F\'isica, Universidad Nacional de Colombia, Carrera 45 $N^o$ 26-85, Bogot\'a D.C., C.P. 111321, Colombia}
\address[b]{Department of Physics, North Carolina State University, Raleigh, NC 27695, USA}

\begin{abstract}
We discuss the non-equilibrium attractors of systems undergoing Gubser flow within kinetic theory by means of nonlinear dynamical systems. We obtain the attractors of anisotropic hydrodynamics, Israel-Stewart (IS) and transient fluid (DNMR) theories. These attractors are non-planar and the basin of attraction is three dimensional. We compare the asymptotic attractors of each hydrodynamic model with the one obtained from the exact Gubser solution of the Boltzmann equation within the relaxation time approximation. Anisotropic hydrodynamics matches, up to high numerical accuracy, the attractor of the exact theory while the other hydrodynamic theories fail to do so. Thus, anisotropic hydrodynamics is an effective theory for far-from-equilibrium fluids, which consists of the dissipative (nonperturbative) contributions at any order in the gradient expansion.
\end{abstract}


\end{frontmatter}


\section*{Introduction}
%
Recent theoretical and phenomenological studies point to the existence of a new theory for far-from-equilibrium fluid dynamics (see Ref.~\cite{Florkowski:2017olj,Romatschke:2017vte,Romatschke:2017ejr} and references therein). In this contribution, we review some of the main results of our research article about non-equilibrium attractor of the Gubser flow in different hydrodynamic schemes~\cite{Behtash:2017wqg}. \\

The Gubser flow describes a conformal system which expands in an azimuthally symmetric way in the transverse plane, and has boost-invariant longitudinal expansion~\cite{Gubser:2010ze}. The symmetries of this flow are manifest in the three dimensional de Sitter spacetime times a line $\ds$~\cite{Gubser:2010ze}. The hydrodynamic approaches used here differ mainly on the shape of the leading background distribution function in momentum space and their particular truncation scheme. The second order hydrodynamic theories (IS and DNMR) expand around a thermal isotropic background while anisotropic hydrodynamics (aHydro) uses a non-equilibrium background which is not rotationally invariant in momentum space.\\

\section*{Results and discussion}
%
\begin{figure}
\centering
\begin{subfigure}[t]{.48\textwidth}
  \centering
  \includegraphics[width=1.0\linewidth]{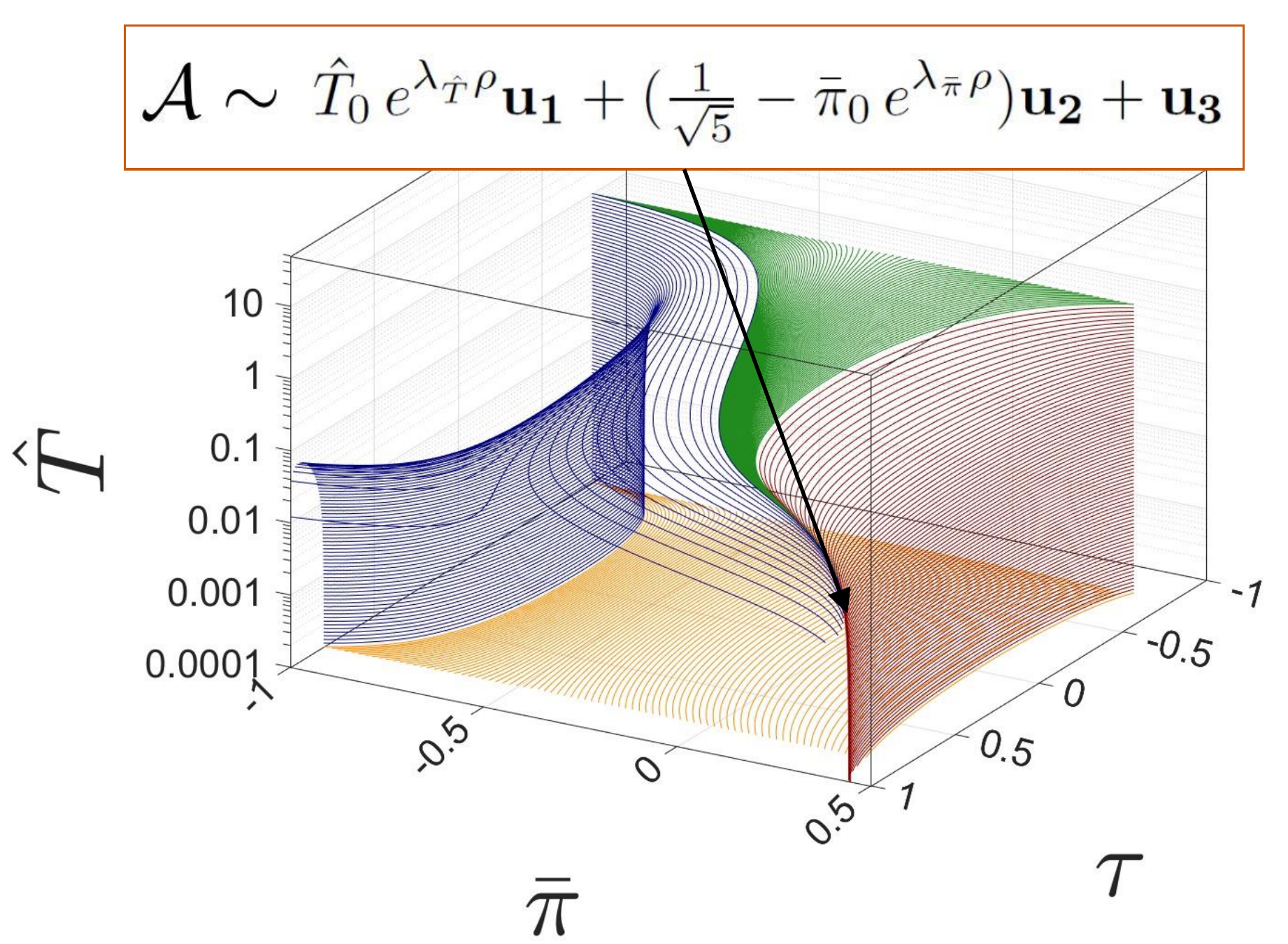}
  \caption{}
  \label{fig:IS}
\end{subfigure}\quad
\begin{subfigure}[t]{.48\textwidth}
  \centering
  \includegraphics[width=1.1\linewidth]{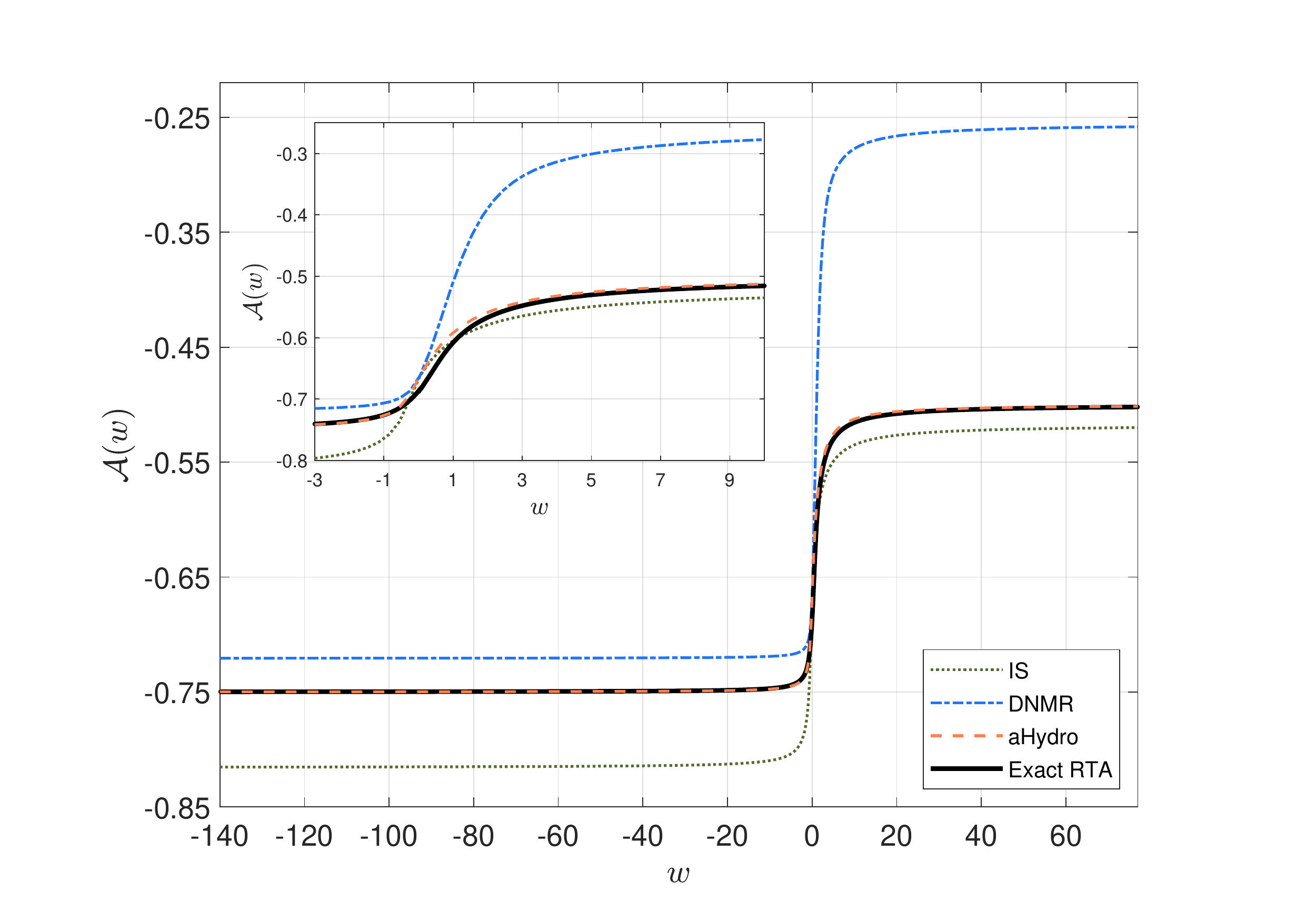}
  \caption{}
  \label{fig:Att}
\end{subfigure}
\caption{(a) $3d$ phase space flows of the IS theory at $\tau_0=-0.8$. We used as initial conditions a rectangular boundary in variables $(\hat{T},\bar{\pi})=\left[1\times10^{-4},\,1\right]\times\left[-1.0,\,0.4\right]$, and let the $3d$ dynamical system evolve. There are two different behaviors observed for the set of studied initial conditions; one for when the flows start inside the basin of attraction, in which case the system evolves to the stable fixed point asymptotically, and the other when the initial condition triggers the flows outside the basin of attraction, in which case the flow lines diverge. The separating blue flow lines determine a portion of the boundary of the basin of attraction. (b) Asymptotic attractors of the IS (dotted green lines), DNMR (dash-dotted blue lines), aHydro (short-dashed orange lines) and exact Gubser solution (solid black line).}
\label{fig:res}
\end{figure}
In Ref.~\cite{Behtash:2017wqg} we showed that the fluid dynamical equations of motion of the macroscopic fields, the temperature $\tem$ and the normalized shear stress $\bar{\pi}$, can be reduced to the following differential equation 
\be
\label{eq:genericfw}
3w\left(\coth^2\rho-1-\mathcal{A}(w)\right)\,\frac{d\mathcal{A}(w)}{dw} +H(\mathcal{A}(w),w) =0\,,
\ee
where $w=\tanh(\rho)/\tem$, $\mathcal{A}=d \log(\tem)/d \log(\cosh(\rho))$ and the functional form of $H$ is model dependent~\cite{Behtash:2017wqg}. The solutions of Eq.~\eqref{eq:genericfw} depends on the value of $\rho$ and only asymptotically (e.g. $\rho\rightarrow \pm \infty$), this equation reduces to a one dimensional ODE. Therefore, the non-equilibrium attractors of systems undergoing Gubser flow are $1d$ {\it non-planar} manifolds~\cite{Behtash:2017wqg}. This can be seen explicitly for the IS theory in Fig.~\ref{fig:IS}, where we plot the flow lines of the fluid dynamical equations in the $3d$ phase space $(\tem,\bar{\pi},\tau=\tanh(\rho))$ for a given set of initial conditions~\cite{Behtash:2017wqg}. In the same figure, we also observe that every set of initial conditions taken from the basin of attraction merges asymptotically to a line, the asymptotic attractor. The rate at which any flow line reaches exponentially the asymptotic attractor is determined by the Lyapunov exponents of the stable fixed point in the dynamical system. A linear stability analysis of the IS theory~\cite{Behtash:2017wqg} reveales that the Lyapunov exponents are $\lambda_{\tem}= -\tfrac{2}{3 } +\tfrac{1}{3\sqrt{5}}$ , $\lambda_{\bar{\pi}}= -\tfrac{8}{3 \sqrt{5}}$ and $\lambda_{\tau}=-1$ around the stable fixed point when $\rho\rightarrow \infty$. This result indicates that in general, neither the Knudsen nor the inverse Reynolds numbers, both proportional to $w$, characterize the asymptotic expansion of the hydrodynamic fields~\cite{Behtash:2017wqg}.\\

In Fig.~\ref{fig:Att} we present the numerical analysis of the asymptotic attractor obtained from Eq.~\eqref{eq:genericfw} when $\rho\rightarrow\infty$ for the IS (dotted green lines), DNMR (dash-dotted blue lines), aHydro (short-dashed orange lines), and the solution to the exact Boltzmann equation for the Gubser flow (solid black line). First, we observe that none of the second order hydrodynamic schemes - DNMR and IS - are able to be in good agreement with the attractor of the exact theory over the large (or small) $w$ regime studied here. Among these two truncation schemes, the DNMR approach is the most unaccurate description of the asymptotic attractor while the IS does a better job. Nonetheless, the IS theory also fails to be valid in the attracting region close to the fixed point. Now, aHydro is the best hydrodynamic scenario for reproducing numerically the attractor. Nonetheless, none of the hydrodynamic models, as expected, are  able to match the exact result around $w\sim 0$ (more precisely between $|w|\lesssim 2$ where $\coth^2\rho-1$ cannot be approximated to zero) as shown in the inset of Fig.~\ref{fig:Att}. Note that the numerical difference between IS and aHydro with respect to the exact result is no larger than $4\%$ in this interval whereas DNMR deviates entirely in there. In the large or intermediate regime, on the other hand, we verify numerically that the largest numerical deviation between the aHydro attractor and the exact one does not exceed $0.06\%$. The numerical results presented here provide a conclusive proof that aHydro resums effectively the Knudsen and inverse Reynolds numbers to all orders independent of the initial conditions.

\section*{Conclusions}
In this proceedings contribution, we analyzed and extended the previous studies of hydrodynamical attractors by taking advantage of powerful and useful tools of nonlinear dynamical systems. These techniques allow us to understand the impossibility of reducing the hydrodynamic evolution equations into a single one as opposed to the Bjorken flow~\cite{Heller:2015dha}. This in turn insicates that the dimension of the basin of attraction is three and one has to consider a genuine 3d initial value problem. \\

From the numerical comparison of the out-of-equilibrium attractors, we conclude that the attractor of the exact theory is best achieved  by aHydro up to high numerical accuracy. The nonperturbative expansion leading to the attractor is well understood in the form of a formal transseries around the hydrodynamic fixed point~\cite{Behtash:2017wqg}. The transseries manifests the sum over all the dissipative (nonperturbative) corrections characterized by the Knudsen and inverse Reynolds numbers at any order in the gradient expansion. \\

\section*{Acknowledgements}

M. M. is supported in part by the US Department of Energy Grant No. DE-FG02-03ER41260 and by the BEST (Beam Energy Scan Theory) DOE Topical Collaboration.


\vspace*{-0.3cm}
\bibliographystyle{elsarticle-num}
\bibliography{nonequilattractor_clean}







\end{document}